\documentclass[a4paper]{jpconf}	

\usepackage{braket}
\usepackage{mathtools,amsmath,amsfonts,amssymb}
\usepackage{accents}

\newcommand{\ul}[1]{\underaccent{\bar}{#1}}
\newcommand{\uc}[1]{\underaccent{\circ}{#1}}
\newcommand{\ub}[1]{\underaccent{\bullet}{#1}}

\usepackage{footmisc}
\DefineFNsymbols*{customfoots}{*{$^\dag$}}
\setfnsymbol{customfoots}

\usepackage{textpos}
\usepackage{etoolbox}
\patchcmd{\thebibliography}{\advance\leftmargin\labelsep}
  {\labelsep=0.5cm \advance\leftmargin\labelsep}{}{}

\begin{document}


\title{Working directly with probabilities in\\ quantum field theory}
\author{R Dickinson$^1$, J Forshaw$^1$ and P Millington$^2$\footnote{Talk presented by P Millington.}}
\address{$^1$Consortium for Fundamental Physics,
  School of Physics and Astronomy,\\
  University of Manchester,
  Manchester M13 9PL,
  United Kingdom}
\address{$^2$School of Physics and Astronomy,\\
  University of Nottingham,
  Nottingham NG7 2RD,
  United Kingdom}
\ead{robert.dickinson-2@manchester.ac.uk,
jeff.forshaw@manchester.ac.uk,
p.millington@nottingham.ac.uk}

\begin{textblock}{4}(12,-9.2)
\begin{flushright}
\begin{footnotesize}
MAN/HEP/2017/04 \\
15 February 2017
\end{footnotesize}
\end{flushright}
\end{textblock}


\begin{abstract}
We present a novel approach to computing transition probabilities in quantum field theory, which allows them to be written directly in terms of expectation values of nested commutators and anti-commutators of field operators, rather than squared matrix elements. We show that this leads to a diagrammatic expansion in which the retarded propagator plays a dominant role. As a result, one is able to see clearly how faster-than-light signalling is prevented between sources and detectors. Finally, we comment on potential implications of this approach for dealing with infra-red divergences.
\end{abstract}

\section{Introduction}

The utility of scattering-matrix theory has biased the development of techniques in quantum field theory towards the calculation of transition \emph{amplitudes}.  However, isolated transition amplitudes are not physical observables and they often suffer from artefacts that are eliminated only after we combine individual amplitudes into transition \emph{probabilities}. For instance, the soft and collinear infra-red (IR) divergences occurring in gauge amplitudes are cancelled by the interference of virtual and real emissions by means of the Bloch-Nordsieck~\cite{Bloch:1937pw} and Kinoshita-Lee-Nauenberg theorems~\cite{Kinoshita:1962ur,Lee:1964is} (in the case of spin-1) or the Weinberg soft graviton theorem~\cite{Weinberg:1965nx} (in the case of spin-2). In non-Abelian gauge theories, unphysical polarisation states are removed by the contributions from Faddeev-Popov ghosts~\cite{Faddeev:1967fc}. It seems reasonable therefore to pursue means of directly calculating transition probabilities that bypass the amplitude level altogether, the hope being that such artefacts never appear explicitly.

Another (not unrelated) artefact of dealing with amplitudes, and one that we will focus on in this note, is apparent a-causal behaviour, i.e.~behaviour seemingly at odds with Einstein causality and the forbiddance of faster-than-light signalling. The archetypal example of such a signalling process is the famous Fermi two-atom problem~\cite{Fermi}. Fermi considered two point-like atoms, A and B, separated by a distance $R$. Atom A is initially prepared in an excited state, and atom B is initially prepared in its ground state. Fermi calculated the probability that, after a time $T$, atom A would be found in its ground state, and atom B would be found in an excited state, following the exchange of a photon. Fermi claimed to prove that this probability is strictly zero for all times $T<R/c$. Fermi was in fact wrong, and one should expect a non-zero probability \emph{for all} $T$, since the simultaneous measurement of the states of atoms A and B does not constitute a local measurement. Instead, we should ask for the probability that atom B is found in an excited state after a time $T$ without making any restriction on the states of atom A or the electromagnetic field~\cite{Shirokov,Ferreti,PT}. (A recent introduction to the Fermi problem and an overview of its history can be found in Ref.~\cite{Dickinson:2016oiy}.) As we will see, this local measurement is independent of the state of atom A for all times $T<R/c$, therefore being manifestly causal in the weak sense (see, e.g.,~Ref~\cite{Hegerfeldt}).

\begin{figure}
\centering
\includegraphics[scale=0.3]{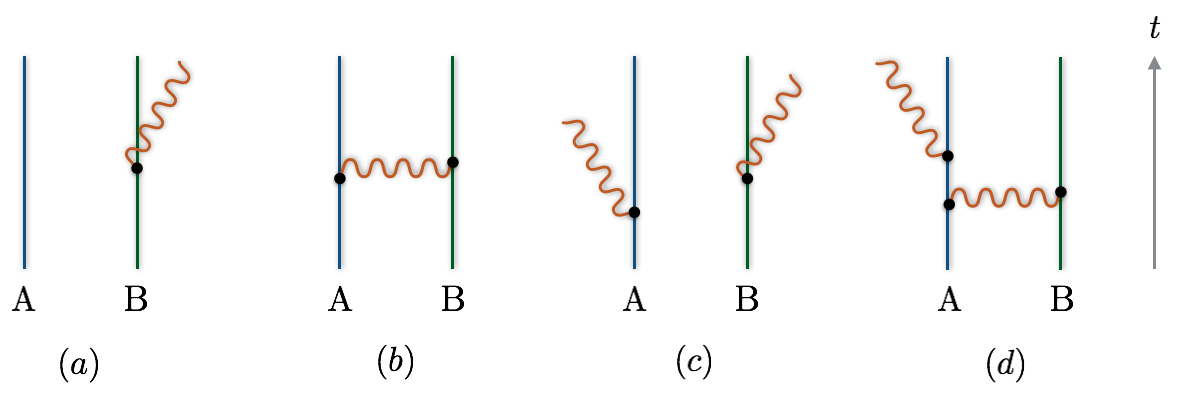}
\caption{\label{fig:Fermi} The four amplitudes relevant to the Fermi problem. Fermi considered diagram ($b$) in isolation, and it can be shown that $|(b)|^2\neq0$ for $T<R/c$. It is only in the combination $|(b)|^2+|(c)|^2+2\,\mathrm{Re}[(a)(d)^*]$ that the a-causal parts cancel. Notice that this cancellation relies on the interference of diagrams involving only \emph{unobserved} emissions.}
\end{figure}

The local measurement described above is fully inclusive over the states of atom A and the electromagnetic field. As such, we must account for unobserved emission of photons in the final state, and the four amplitudes relevant to the Fermi problem are shown in Fig.~\ref{fig:Fermi}. By working directly at the level of the probability, we will find that we do not need to sum explicitly over these unobserved emissions~\cite{Dickinson:2016oiy}. This suggests that one may be able to write down semi-inclusive transition probabilities in which the Bloch-Nordsieck cancellation is applied implicitly, potentially having a significant impact on the way in which we deal with IR divergences in gauge theories. As a step towards this, we will conclude the present note by describing an example of a Fock-space projection operator for (semi-)inclusive observables (see Ref.~\cite{Dickinson2017}).

\section{Causality}

Causality is built into quantum field theory \emph{ab initio} through the vanishing of the equal-time commutator of field operators, in the case of bosons, or anti-commutator of field operators, in the case of fermions. This is the so-called \emph{microcausality condition}. For example, given a real scalar field $\phi_x\equiv\phi(x)$, we have
\begin{equation}
\big[\phi_x,\:\phi_y\big]\ =\ 	0\qquad \forall\ (x-y)^2\ <\ 0\qquad \text{(space-like)}\;.
\end{equation}
However, it is the Feynman propagator
\begin{equation}
\label{eq:Feynman}
\Delta^{({\rm F})}_{xy}\ \equiv\ \Delta^{({\rm F})}(x,y)\ =\ \frac{1}{2}\,\mathrm{sgn}(x_0-y_0)\braket{\big[\phi_x,\:\phi_y\big]}\:+\:\frac{1}{2}\,\braket{\big\{\phi_x,\:\phi_y\big\}}\ =\ \int\!\frac{{\rm d}^4k}{(2\pi)^4}\,\frac{ie^{-ik\cdot (x-y)}}{k_0^2-E_{\mathbf{k}}^2+i\epsilon}
\end{equation}
that is ubiquitous in $S$-matrix theory and it has support over space-like separations due to the anti-commutator term $\smash{\braket{\big\{\phi_x,\:\phi_y\big\}}}$. In Eq.~\eqref{eq:Feynman}, \smash{$E_{\mathbf{k}}=\sqrt{\mathbf{k}^2+m^2}$} is the on-shell energy (for a field of mass $m$), and $\smash{\mathrm{sgn}(x_0-y_0)}$ is the signum function, which may be expressed in terms of unit-step functions as $\mathrm{sgn}(x_0-y_0)\equiv\Theta(x_0-y_0)-\Theta(y_0-x_0)$.

The fact that causality is not manifest at the amplitude level can be understood directly in terms of the Feynman prescription ($k_0^2\to k_0^2+i\epsilon$, $\epsilon=0^+$), since this choice of analytic continuation allows us to Wick rotate between Minkowski space-time, which has causal structure, and four-dimensional Euclidean space, which does not. It is perhaps even more disconcerting that the a-causal part of the Feynman propagator actually corresponds to \emph{on-shell} propagation, that is the part that we would normally associate with \emph{real} particles. If causality is to be manifest then it is surely the retarded and advanced propagators
\begin{equation}
\Delta_{xy}^{({\rm R})}\ = \ \Delta_{yx}^{({\rm A})}\ =\ \frac{1}{i}\,\Theta(x_0-y_0)\braket{\big[\phi_x,\:\phi_y\big]}\ =\ \int\!\frac{{\rm d}^4 k}{(2\pi)^4}\,\frac{e^{-ik\cdot(x-y)}}{k_0^2-E_{\mathbf{k}}^2\pm i\epsilon\,\mathrm{sgn}(k_0)}
\end{equation}
that should play the dominant role, since these have support only over light- and time-like separations.  Note that neither the retarded ($k_0^2\to k_0^2+i\epsilon\,\mathrm{sgn}(k_0)$, $\epsilon=0^+$) nor advanced ($k_0^2\to k_0^2-i\epsilon\,\mathrm{sgn}(k_0)$, $\epsilon=0^+$) pole prescriptions permit us to Wick rotate, as we should expect. In the next section, we will show that the retarded and advanced propagators do in fact play this dominant role but only at the level of \emph{probabilities}, and their presence will ensure that faster-than-light signalling is forbidden. Moreover, we will see that they emerge through the interplay of time- and anti-time-ordered products of field operators.

Time-ordering of the two-point function corresponds to the Feynman prescription ($k_0^2 \to k_0^2+i\epsilon$), which places the positive- and negative-frequency poles in the second and fourth quadrants of the complex plane, respectively. Instead, anti-time-ordering corresponds to the Dyson prescription ($k_0^2 \to k_0^2-i\epsilon$), which places the positive- and negative-frequency poles in the first and third quadrants, respectively. It is through combinations of terms involving these two prescriptions that both poles can appear in the lower-half complex plane to give the retarded propagator or the upper-half complex plane to give the advanced propagator.

The role of both time orderings can also be illustrated by means of the Bogoliubov-Shirkov causality condition~\cite{BogShirk}.  We begin with the familiar $S$-matrix operator
\begin{equation}
\label{eq:Smatrix}
S[\lambda]\ =\ \mathrm{T}\exp\bigg[-i\!\int\!{\rm d}^4 z\,\mathcal{H}^{\rm int}_z(\lambda)\bigg]\;,
\end{equation}
where $\mathrm{T}$ is the time-ordering operator and the domain of integration over $z_0$ is $]-\infty,\infty[$. Taking a local interaction Hamiltonian density $\mathcal{H}^{\rm int}_z(\lambda)=\lambda\,\phi^4_z/4!$, we imagine promoting the coupling $\lambda$ to one of two space-time dependent functions $\lambda_z\equiv\lambda(z)$ or $\lambda'_z\equiv\lambda'(z)$. If the coupling functions $\lambda_z$ and $\lambda_z'$ coincide for all $z_0$ earlier than some time $t$, the product $S[\lambda']S^{\dag}[\lambda]$ must be independent of the behaviour of these functions for all times $z_0<t$ by unitarity. Letting $\lambda'_z=\lambda_z+\delta\lambda_z$, where the infinitesimal variation $\delta\lambda_z$ is only non-zero for $z_0>t$, we expand
\begin{equation}
\label{eq:SSdag}
S[\lambda']S^{\dag}[\lambda]\ =\ S[\lambda]S^{\dag}[\lambda]\:+\:\frac{\delta S[\lambda]}{\delta \lambda_y}\,\delta\lambda_y\,S^{\dag}[\lambda]\ =\ \mathbb{I}\:+\:\frac{\delta S[\lambda]}{\delta \lambda_y}\,\delta\lambda_y\,S^{\dag}[\lambda]\;,
\end{equation}
in which the integral over $y$ (for $y_0>t$) in the first functional variations is left implicit. If this product is to be consistent with causality, Eq.~\eqref{eq:SSdag} must still be independent of the behaviour of the coupling function for all times $x_0<t<y_0$ and, by Lorentz covariance, for all $x$ and $y$ that are space-like separated. We therefore require that the quantity
\begin{equation}
\frac{\delta}{\delta \lambda_x}\bigg[\frac{\delta S[\lambda]}{\delta \lambda_y}\,S^{\dag}[\lambda]\bigg]
\end{equation}
vanish for all $x\precsim y$, i.e.~for all $x$ causally preceding or space-like separated from $y$. This gives the Bogoliubov-Shirkov causality condition~\cite{BogShirk}:
\begin{equation}
\label{eq:BogShirk}
\frac{\delta^2S[\lambda]}{\delta \lambda_x\,\delta \lambda_y}\,S^{\dag}[\lambda]\:+\:\frac{\delta S[\lambda]}{\delta \lambda_y}\,\frac{\delta S^{\dag}[\lambda]}{\delta \lambda_x}\ =\ 0 \qquad \forall\ x\ \precsim\ y\;,
\end{equation}
which can be satisfied only through the cancellation of terms originating from $S$ (time-ordered) and $S^{\dag}$ (anti-time-ordered).

Making use of the explicit form of the $S$-matrix operator in Eq.~\eqref{eq:Smatrix}, Eq.~\eqref{eq:BogShirk} can be written
\begin{equation}
\mathrm{T}\bigg[\frac{\partial \mathcal{H}^{\rm int}_x}{\partial \lambda_x}\,\frac{\partial \mathcal{H}^{\rm int}_y}{\partial \lambda_y}\,S\bigg]S^{\dag}\:-\:\mathrm{T}\bigg[\frac{\partial \mathcal{H}^{\rm int}_y}{\partial \lambda_y}\,S\bigg]\bar{\mathrm{T}}\bigg[\frac{\partial \mathcal{H}^{\rm int}_x}{\partial \lambda_x}\,S^{\dag}\bigg]\ =\ 0 \qquad \forall\ x\ \precsim\ y\;,
\end{equation}
where $\bar{\mathrm{T}}$ is the anti-time-ordering operator. At lowest order in the coupling function, this requires
\begin{equation}
\label{eq:Scomm}
\Theta(x_0-y_0)\bigg[\frac{\partial \mathcal{H}^{\rm int}_x}{\partial \lambda_x},\: \frac{\partial \mathcal{H}^{\rm int}_y}{\partial \lambda_y} \bigg]\ =\ 0 \qquad \forall\ x\ \precsim\ y\;,
\end{equation}
and, on substituting for the interaction Hamiltonian density, we obtain the constraint
\begin{equation}
\label{eq:largesttime}
\frac{i}{(3!)^2}\,\Delta_{xy}^{({\rm R})}\phi_x^3\phi_y^3\:+\:\frac{1}{(2!)^3}\,[\Delta_{xy}^{({\rm R})}]^2\phi_x^2\phi_y^2\:-\:\frac{i}{3!}\,[\Delta_{xy}^{({\rm R})}]^3\phi_x\phi_y\:-\:\frac{1}{4!}[\Delta_{xy}^{({\rm R})}]^4\ =\ 0 \qquad \forall\ x\ \precsim\ y\;.
\end{equation}
This is automatically satisfied due to the presence of the retarded propagators,\footnote{We have made use of the result (valid for all integer $n,m\geq1$)
\begin{equation*}
\big[\phi_x^n,\:\phi_y^m\big]\ =\ \sum_{k\,=\,1}^{\mathrm{min}(n,m)}(-1)^{k+1}\,\frac{n! m!}{k!(n-k)!(m-k)!}\,\big[\phi_x,\:\phi_y\big]^k\phi_x^{n-k}\phi_y^{m-k}\;.
\end{equation*}}
which have originated from the commutator in Eq.~\eqref{eq:Scomm}. Proceeding to higher orders in the coupling, the combination of time- and anti-time-ordered products will give rise to a series of nested commutators of the Hamiltonian density that is again consistent with causality. In the next section, we will show that this series can be obtained straightforwardly at the level of expectation values by application of the Baker-Campbell-Hausdorff formula.

We remark that the Bogoliubov-Shirkov causality condition [Eq.~\eqref{eq:BogShirk}] can be expressed in terms of the largest time equation~\cite{tHooft:1973wag}, of which Eq.~\eqref{eq:largesttime} is an example. This hints at the fundamental  relationship between unitarity and causality. This connection has recently been emphasised~\cite{Tomboulis:2017rvd} by means of the tree-loop duality relations~\cite{Catani:2008xa,Bierenbaum:2010cy}, and it can be made explicit (see Ref.~\cite{Dickinson:2013lsa}) through the Kobes-Semenoff unitarity cutting rules~\cite{Kobes:1985kc,Kobes:1986za,Kobes:1990ua} of the Schwinger-Keldysh closed-time-path (or in-in) formalism~\cite{Schwinger:1960qe,Keldysh:1964ud}. 


\section{Sources and detectors}
\label{sec:SD}

We now turn out attention to the signalling between sources and detectors. For simplicity, we introduce a scalar analogue of the Fermi two-atom problem~\cite{Dickinson:2016oiy}, which consists of a real scalar field $\phi$ that interacts with two static ``atoms'' $S$ and $D$, fixed at positions $\mathbf{x}^S$ and $\mathbf{x}^D$, respectively. Working in the interaction picture, the free part of the Hamiltonian is
\begin{equation}
H_0 \ = \  \sum_n \omega_n^S \ket{n^S}\!\bra{n^S}\: +\: \sum_n \omega_n^D \ket{n^D}\!\bra{n^D}\: +\: \int {\rm d}^3\mathbf{x}\; \Big( \tfrac{1}{2}(\partial_t\phi)^2\: +\: \tfrac{1}{2}(\mathbf{\nabla}\phi)^2\: +\: \tfrac{1}{2}m^2\phi^2 \Big)\;,
\end{equation}
where the states $\{\ket{n^X}\}$ form a complete set of bound states for atom $X\in\{S,D\}$. The interaction Hamiltonian is
\begin{subequations}
\begin{gather}
H_{\mathrm{int}}(t) \ = \ M^S(t)\,\phi(\mathbf{x}^S,t)\:+\: M^D(t)\,\phi(\mathbf{x}^D,t)\;,\\ M^X(t)\ \equiv\ \sum_{m\,\neq\, n}\mu_{mn}^X\,e^{i\omega_{mn}^X t}\ket{m^X}\!\bra{n^X}\;,
\end{gather}
\end{subequations}
where $\omega_{mn}^X\equiv\omega_{m}^X-\omega_n^X$ and $\mu_{mn}^X$ are monopole moments.

At time $t=0$, the system is prepared with atom $D$ in its ground state ($\ket{g^D}$), the field $\phi$ in its vacuum state ($\ket{0^{\phi}}$) and atom $S$ in an incoherent superposition of its ground and excited states ($\ket{g^S}$ and $\ket{p^S}$). This corresponds to an initial density operator
\begin{equation}
\rho_0\ \equiv\ \rho(0)\ =\ \gamma\ket{i_p}\!\bra{i_p}\:+\:(1-\gamma)\ket{i_g}\!\bra{i_g}\;,\qquad \ket{i_{p(g)}}\ =\ \ket{p(g)^{S}}\:\otimes\:\ket{g^D}\:\otimes\:\ket{0^{\phi}}\;,
\end{equation}
where $\gamma\in\mathbb{R}$ parametrises the admixture of the initial states of the source atom $S$. The measurement (at some later time $t=T$) is effected by an operator $E$, which projects into the subspace of states in which the detector atom $D$ is in its excited state ($\ket{q^D}$) but with no restriction on the states of atom $S$ or the field $\phi$:
\begin{equation}
\label{eq:effop}
E\ \equiv\ E^S\:\otimes\:E^D\:\otimes\:\mathcal{E}\ =\ \mathbb{I}^S\:\otimes\:\ket{q^D}\!\bra{q^D}\:\otimes\:\mathbb{I}^{\phi}\;.
\end{equation}
The probability of the local measurement outcome ``atom $D$ excited'' is then
\begin{equation}
\label{eq:mout}
\mathbb{P}(D^*) \ =\  \mathrm{Tr}\,(E \rho_T)\ =\ \mathrm{Tr}\,(U_{T,0}^{\dag}EU_{T,0}\rho_0)\;,
\end{equation}
where
\begin{equation}
U_{T,0}\ =\ \mathrm{T}\exp\bigg[-i\!\int_0^T\!{\rm d}t\;H_{\rm int}(t)\bigg]
\end{equation}
is the time-evolution operator. It is important to draw a clear distinction between the probability of the measurement outcome in Eq.~\eqref{eq:mout} and the sensitivity of the detector to the initial preparation of the system. The latter tells us about our ability to signal between the source and detector atoms, and it is given by
\begin{equation}
\label{eq:sigma}
\sigma_{pg}\ \equiv\ \frac{{\rm d}\,\mathbb{P}(D^*)}{{\rm d}\gamma}\ =\ \mathbb{P}_p(D^*)\:-\:\mathbb{P}_g(D^*)\;,\qquad \mathbb{P}_{p(g)}(D^*)\ \equiv\ \braket{i_{p(g)}|U_{T,0}^{\dag}EU_{T,0}|i_{p(g)}}\;.
\end{equation}

By applying the Baker-Campbell-Hausdorff formula to Eq.~\eqref{eq:sigma}, the combination of time-ordered (from $U$) and anti-time-ordered exponentials (from $U^{\dag}$) gives rise to an infinite series of nested commutators~\cite{Dickinson:2016oiy,Dickinson:2013lsa,Cliche:2009fma,FransonDonegan}. This can be written in the form~\cite{Dickinson:2016oiy,Dickinson:2013lsa}
\begin{equation}
\label{eq:Ppg}
\mathbb{P}_{p,g}\ =\ \sum_{j\,=\,0}^\infty \int_0^{T} {\rm d}t_1\,{\rm d} t_2\ldots {\rm d} t_j \;\Theta_{12\dots j} \bra{i_{p,g}}\mathcal{F}_j\ket{i_{p,g}} \;,
\end{equation}
where $\mathcal{F}_0  \equiv  E$,
\begin{equation}
\mathcal{F}_j \ =\  \tfrac{1}{i}\Big[ \mathcal{F}_{j-1},\: H_\mathrm{int}(t_j) \Big]\ =\ \tfrac{1}{i}\Big[ \mathcal{F}_{j-1},\: M^S_j\phi^S_j\:+\: M^D_j\phi^D_j \Big]\;,
\end{equation}
and $\Theta_{ijk\dots}$ is equal to $1$ for $t_i>t_j>t_k>\dots$ and $0$ otherwise. We use the short-hand notation $M_j^X\equiv M^X(t_j)$ and $\phi^X_j\equiv\phi(\mathbf{x}^X,t_j)$.  By defining
\begin{subequations}
\begin{gather}
E^X_{\dots k}\ \equiv\ \tfrac{1}{i}\big[E^X_{\dots},\:M^X_k\big]\;,\qquad E^X_{\dots \ul{k}}\ \equiv\ \big\{E^X_{\dots},\:M_k^X\big\}\;,\\
\mathcal{E}^{\dots X}_{\dots k}\ \equiv\ \tfrac{1}{i}\big[\mathcal{E}^{\dots}_{\dots},\:\phi^X_k\big]\;,\qquad \mathcal{E}^{\dots X}_{\dots \ul{k}}\ \equiv\ \big\{\mathcal{E}^{\dots}_{\dots},\:\phi^X_k\big\}\;,
\end{gather}
\end{subequations}
we can introduce a convenient ``undercircle'' notation~\cite{Dickinson:2016oiy}
\begin{equation}
E_{\uc{k}\uc{l}}\mathcal{E}_{\ub{k}\ub{l}}\ \equiv\ E_{kl}\mathcal{E}_{\ul{k}\ul{l}}\:+\: E_{k\ul{l}}\mathcal{E}_{\ul{k}l}\:+\: E_{\ul{k}l}\mathcal{E}_{k\ul{l}}\:+\: E_{\ul{k}\ul{l}}\mathcal{E}_{kl}\;,
\end{equation}
which allows us to write the $n$-th order operator appearing in Eq.~\eqref{eq:Ppg} as
\begin{equation}
\label{eq:Fn}
\mathcal{F}_n \ =\  2^{-n}\sum_{a\,=\,0}^{n}E^S_{(\uc{1}\ldots\!\underaccent{\!\!\cdots}{}\; \uc{a}}\,E^D_{a+\!\uc{}\,1\ldots\!\!\!\underaccent{\cdots}{}\;\;\uc{n})}\,\mathcal{E}^{(\!S\ldots S\,\ D\ldots D)}_{(\ub{1}\,\ldots\underaccent{\!\!\!\!\cdots}{}\,\ub{a}\,a+\!\ub{}\,1\ldots\!\!\!\underaccent{\cdots}{}\;\;\ub{n})}\;.
\end{equation}
The parentheses indicate a summation over all unique ordered permutations of the indices, e.g., $E^S_{(12}E^D_{3)}=E_{12}^SE_3^D+E_{13}^SE_2^D+E^S_{23}E_1^D$.

In the effect operator [Eq.~\eqref{eq:effop}], we have made no restriction on the states of atom $S$ or the field $\phi$, fixing $E^S=\mathbb{I}^S$ and $\mathcal{E}=\mathbb{I}^{\phi}$. We may then show that:
\begin{itemize}
\item [(i)] $E_{k\dots}^S\ =\ 0$, such that the first index on $E^S_{\dots}$ must always be underlined, with $E_{\ul{k}}^S=2M_k^S$.

\item [(ii)] $\mathcal{E}_{1\dots}^{X\dots}=0$ and $\mathcal{E}_{\ul{1}23\dots}^{XYZ\dots}=0$, such that the `1' index (the latest time) is never underlined on $E_{1\dots}^X$ and, more generally, any $\mathcal{E}^{\dots}_{\dots}$ operator vanishes when its first $k$ indices consist of more non-underlined than underlined indices for any $k$.
\end{itemize}
These restrictions allow us to eliminate a large proportion of the terms in Eq.~\eqref{eq:Fn}.

\begin{figure}
\centering
\includegraphics[scale=0.3]{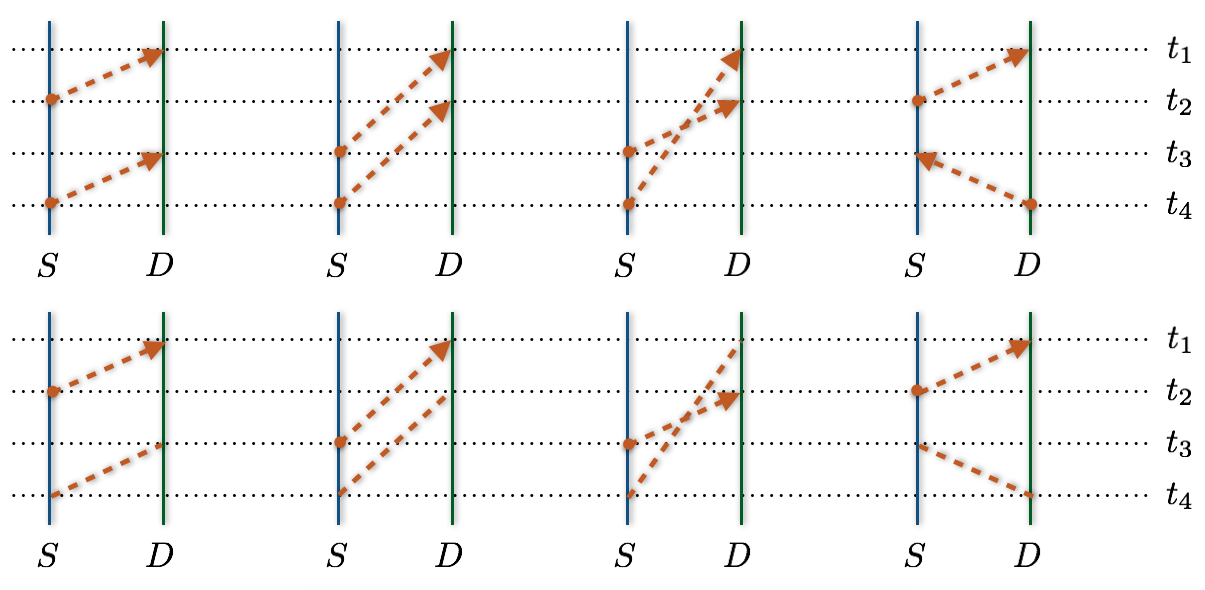}
\caption{\label{fig:keyresult} Diagrammatic representation of $\mathcal{F}_4$, describing the evolution of the density operator forwards in time due to the exchange of a $\phi$ quantum. The initial time $t=0$ is at the bottom of the diagrams and the time of the measurement $t=T$ is at the top of the diagrams. The solid vertical lines track the source (blue) and detector (green) atoms.  Orange dashed lines with arrows are associated with retarded propagators {\footnotesize $\Delta_{ij}^{XY({\rm R})}$}, and those without arrows are associated with Hadamard propagators {\footnotesize $\Delta_{ij}^{XY({\rm H})}$}. Each vertex is associated with a monopole moment $\mu_{mn}^X$.}
\end{figure}

The first non-vanishing contributions to Eq.~\eqref{eq:sigma} arise from the fourth-order operator
\begin{align}
\mathcal{F}_4 \ &=\  \tfrac{1}{16}\big(
E^D_{1{2}\uc{3}\uc{4}}\mathcal{E}^{D\!D\!D\!D}_{\ul{1}\ul{2}\ub{3}\ub{4}}
\:+\: E^D_{1\ul{2}{3}\uc{4}}\mathcal{E}^{D\!D\!D\!D}_{\ul{1}{2}\ul{3}\ub{4}}
\:+\: E^D_{1{2}\uc{3}}E^S_{\ul{4}}\mathcal{E}^{D\!D\!D\!S}_{\ul{1}\ul{2}\ub{3}4} 
\:+\: E^D_{1\ul{2}{3}}E^S_{\ul{4}}\mathcal{E}^{D\!D\!D\!S}_{\ul{1}{2}\ul{3}4} \nonumber\\&\;\;\;\;\;\;\;\;\;\;
+\: E^D_{1{2}\uc{4}}E^S_{\ul{3}}\mathcal{E}^{D\!D\!S\!D}_{\ul{1}\ul{2}3\ub{4}} 
\:+\: E^D_{1{3}\uc{4}}E^S_{\ul{2}}\mathcal{E}^{D\!S\!D\!D}_{\ul{1}2\ul{3}\ub{4}}  
\:+\: E^D_{12}E^S_{\ul{3}\uc{4}}\mathcal{E}^{D\!D\!S\!S}_{\ul{1}\ul{2}3\ub{4}} 
\:+\: E^D_{1{3}}E^S_{\ul{2}\uc{4}}\mathcal{E}^{D\!S\!D\!S}_{\ul{1}2\ul{3}\ub{4}}\nonumber\\&\;\;\;\;\;\;\;\;\;\;
+\: E^D_{1\uc{4}}E^S_{\ul{2}{3}}\mathcal{E}^{D\!S\!S\!D}_{\ul{1}2\ul{3}\ub{4}} 
\:+\: E^D_{1}E^S_{\ul{2}{3}\uc{4}}\mathcal{E}^{D\!S\!S\!S}_{\ul{1}2\ul{3}\ub{4}} 
\big)\;.
\end{align}
Substituting this expression into Eq.~\eqref{eq:sigma}, the leading-order result for the sensitivity of atom $D$ to the preparation of the system is~\cite{Dickinson:2016oiy}
\begin{align}
\label{eq:keyresult}
\frac{{\rm d}\,\mathbb{P}(D^*)}{{\rm d}\gamma}\ &=\ 2\sum_n\int_{t_1>t_2>t_3>t_4}|\mu_{pn}^S|^2|\mu_{qg}^D|^2\nonumber\\&\qquad\times\:\bigg[\cos(\omega_{qg}^Dt_{12})\Big(\sin(\omega_{pn}^St_{34})\Delta_{24}^{DS({\rm H})}\:+\:\cos(\omega_{pn}^St_{34})\Delta_{24}^{DS({\rm R})}\Big)\Delta_{13}^{DS({\rm R})}\nonumber\\&\qquad \ \ +\:\cos(\omega_{qg}^Dt_{12})\Big(\sin(\omega_{pn}^St_{34})\Delta_{14}^{DS({\rm H})}\:+\:\cos(\omega_{pn}^St_{34})\Delta_{14}^{DS({\rm R})}\Big)\Delta_{23}^{DS({\rm R})}\nonumber\\&\qquad \ \  +\:\cos(\omega_{qg}^Dt_{13})\Big(\sin(\omega_{pn}^St_{24})\Delta_{34}^{DS({\rm H})}\:+\:\cos(\omega_{pn}^St_{24})\Delta_{34}^{DS({\rm R})}\Big)\Delta_{12}^{DS({\rm R})}\nonumber\\&\qquad \ \ +\:\sin(\omega_{pn}^{S}t_{23})\Big(\cos(\omega_{qg}^{D}t_{14})\Delta_{34}^{SD({\rm H})}\:+\:\sin(\omega_{qg}^{D}t_{14})\Delta_{34}^{SD({\rm R})}\Big)\Delta_{12}^{DS({\rm R})}\bigg]\:+\:\cdots\;,
\end{align}
where $t_{ij}\equiv t_i-t_j$, and we have defined the retarded and Hadamard propagators
\begin{subequations}
\begin{gather}
\Delta_{jk}^{XY({\rm R})}\ \equiv\ \tfrac{1}{i}\,\Theta_{jk}\braket{\big[\phi_j^X,\:\phi_k^Y\big]}\;,\\
\Delta_{jk}^{XY({\rm H})}\ \equiv\ \braket{\big\{\phi_j^X,\:\phi_k^Y\big\}}\;.
\end{gather}
\end{subequations}
Most importantly, we see that the latest time on the source atom is always connected to a later time on the detector atom by a retarded propagator. As a result, Eq.~\eqref{eq:keyresult} is strictly zero if the source and detector atoms are space-like separated, and faster-than-light signalling is manifestly prohibited. The expectation value of $\mathcal{F}_4$ is illustrated diagrammatically in Fig.~\ref{fig:keyresult}. A full exposition of the diagrammatic rules for this system is presented in Ref.~\cite{Dickinson:2016oiy}; this includes results for the expectation values of general nestings of commutators and anti-commutators of field operators, as well as the rules for computing the expectation values of the atom operators.

We remark that the above approach can be generalised to include additional fields, as well as their mutual and self-interactions; local interactions that involve spatial and temporal derivatives of the fields; and spatially extended sources and detectors (see Ref.~\cite{Dickinson:2016oiy}). In all of these cases, the manifest causality of local measurements persists.


\section{Implications for (semi-)inclusive observables}

In order to obtain a causal result in Sec.~\ref{sec:SD}, it was necessary to sum inclusively over the unobserved final states of the would-be photon field $\phi$ (and source atom $S$). Doing otherwise would have required us to sample the state of the field everywhere simultaneously, which would not constitute a local measurement. However, if we compare Figs.~\ref{fig:Fermi} and \ref{fig:keyresult}, we notice that the unobserved emissions necessarily accounted for in the amplitude-level calculation never appeared explicitly in the probability-level calculation. This can be traced back to the simple fact that, at the probability level, the effect operator corresponding to a fully inclusive measurement is just the unit operator, which trivially commutes with everything else, dropping out of the calculation entirely. The import of this observation is the following: by working directly with probabilities, we do not have to keep track of and calculate the individual amplitudes for all possible unobserved emissions in the final state.

Notwithstanding the potential implications of this for treating IR divergences, probability-level calculations allow us to introduce projection operators for (semi-)inclusive measurements, which do not have simple analogues at the amplitude level~\cite{Dickinson2017}. As an example, the effect operator that sums inclusively over final-state radiation with momenta below some scale $\mu$ is
\begin{equation}
\label{eq:Sudokov}
E_{\mu}\ =\ :e^{-N_\mu}:\;,\qquad N_{\mu}\ =\ \int\!\frac{{\rm d}^3\mathbf{k}}{(2\pi)^3}\,\frac{1}{2E_{\mathbf{k}}}\;\theta(|\mathbf{k}|-\mu)\,a_{\mathbf{k}}^{\dag}a_{\mathbf{k}}\;.
\end{equation}
The colons indicate normal ordering, and $a^{\dag}_{\mathbf{k}}$ and $a_{\mathbf{k}}$ are the usual scalar creation and annihilation operators. In the limit $\mu\to\infty$, the operator is fully inclusive over final-state radiation:
\begin{subequations}
\label{eq:fullinc}
\begin{gather}
\lim_{\mu\,\to\,\infty}E_{\mu}\ = \ \mathbb{I}\;,\\
\lim_{\mu\,\to\,\infty}\braket{0|\phi(x_1)\phi(x_2)E_{\mu}\phi(x_3)\phi(x_4)|0}\ = \ \braket{0|\phi(x_1)\phi(x_2)\phi(x_3)\phi(x_4)|0}\;,
\end{gather}
\end{subequations}
where the appearance of the four-point amplitude is a consequence of the optical theorem. In the limit $\mu\to0$, the operator projects out the vacuum exclusively (see, e.g.,~Ref.~\cite{Louisell}):
\begin{subequations}
\begin{gather}
\lim_{\mu\,\to\,0}E_{\mu}\ = \ \ket{0}\!\bra{0}\;,\\
\lim_{\mu\,\to\,0}\braket{0|\phi(x_1)\phi(x_2)E_{\mu}\phi(x_3)\phi(x_4)|0}\ = \ \braket{0|\phi(x_1)\phi(x_2)|0}\!\braket{0|\phi(x_3)\phi(x_4)|0}\;.
\end{gather}
\end{subequations}
For finite $\mu$, Eq.~\eqref{eq:Sudokov} defines an operator form of the Sudakov factor~\cite{Sudakov:1954sw}.
Such an object could not be written down at the amplitude level. 


\section{Concluding remarks}

In this note, we have outlined a method for directly calculating transition probabilities in quantum field theory in terms of the expectation values of nested commutators and anti-commutators of field operators. These transition probabilities are manifestly causal and have the interesting feature that unobserved emissions are accounted for \emph{implicitly} in (semi-)inclusive measurements. The ability to sum implicitly over soft emissions may have important implications for how we deal with IR divergences in gauge theories, and this motivates the further development of technologies for evaluating these probabilities. Such technology may, for instance, exploit the connection between causality and unitarity to which we have alluded, making use of the unitarity cutting rules of the closed-time-path formalism and building upon ideas presented in~Ref.~\cite{Dickinson:2013lsa}.


\ack

The work of PM is supported by STFC grant ST/L000393/1. PM would like to thank the organisers and participants of DICE2016 for their hospitality, questions and comments.


\end{document}